# An Algorithm for Computing Stochastically Stable Distributions with Applications to Multiagent Learning in Repeated Games

**John R. Wicks**
Department of Computer Science
Brown University, Box 1910
Providence, RI 02912

**Amy Greenwald**
Department of Computer Science
Brown University, Box 1910
Providence, RI 02912

## Abstract

One of the proposed solutions to the equilibrium selection problem for agents learning in repeated games is obtained via the notion of stochastic stability. Learning algorithms are perturbed so that the Markov chain underlying the learning dynamics is necessarily irreducible and yields a unique stable distribution. The stochastically stable distribution is the limit of these stable distributions as the perturbation rate tends to zero. We present the first exact algorithm for computing the stochastically stable distribution of a Markov chain. We use our algorithm to predict the long-term dynamics of simple learning algorithms in sample repeated games.

## 1 Introduction

Recently, game theorists and AI researchers alike have been addressing the question: what is the outcome of multiagent learning in repeated games? Game-theoretic equilibrium concepts are obvious candidate solutions to this problem. However, most equilibrium concepts (e.g., Nash) give rise to sets of equilibria, rather than unique equilibria, so that learners typically face an equilibrium selection problem.

For example, consider the two-player, two-action game QWERTY, studied in Young (c1998) (see Game 1). This game is a coordination game where the players can coordinate their behavior to choose either DVORAK keyboards or QWERTY keyboards. Payoffs are slightly higher if the players manage to coordinate on DVORAK, but coordinating on QWERTY is far better than not coordinating at all.

Typical learning models, such as best-response dynamics (Cournot, 1838) or fictitious play (Brown, 1951), could converge to either the DVORAK equilibrium or

**Game 1: QWERTY**

|   | d | q |
|---|---|---|
| D | 5,5 | 0,0 |
| Q | 0,0 | 4,4 |

the QWERTY equilibrium in this game. Anecdotally, we report that regret matching (Hart and Mas-Colell, 2000), another popular learning model, converges to the DVORAK equilibrium more often than it converges to the QWERTY equilibrium, but either equilibrium can arise under appropriate conditions.

From a theoretical point of view, it is difficult to make predictions about the outcome of learning in games with multiple equilibria because of the dependencies that arise on the path to any equilibrium (e.g., dependence on initial conditions). An empirical approach to making such predictions might be to first simulate the learning dynamics to convergence say 1000 times, and then to make predictions accordingly.

Alternatively, Foster and Young (1990) argue that players are prone to make mistakes, so it is reasonable to augment standard learning models with mistake probabilities. Under this assumption, learning dynamics never converge, but instead tend to oscillate, usually from one equilibrium to another. With such persistent randomness embedded in learning models, are some outcomes more likely than others?

Young (1993) goes on to apply the theory of stochastic stability to make predictions about the long-term dynamics of multiagent learning in repeated games. Given a learning model, its dynamics can be described by an underlying Markov process. By introducing mistake probabilities into the model, the underlying process becomes a perturbed Markov process. A *perturbed Markov process* is a family of Markov processes indexed by a perturbation rate $\epsilon$: e.g., a mistake probability. Each Markov process $M_\epsilon$ in such a family has a unique stable distribution, call it $v_\epsilon$. The limit of the sequence $\{v_\epsilon\}$ as $\epsilon \to 0$ exists, is unique, and is called the *stochastically stable distribution* of the perturbed Markov process. The stochastically stable distribution can be interpreted as a description of the long-run dynamics of the original learning model.

In particular, the states in the support of the stochastically stable distribution (i.e., those with strictly positive probabilities) are called the *stochastically stable states* of a perturbed Markov process. The probability of a stochastically stable state does not vanish as the amount of noise goes to zero. Young presents an algorithm for computing stochastically stable states of a perturbed Markov process, and applies his algorithm to obtain such results as: the unique outcome in $2 \times 2$ coordination games is the so-called "risk-dominant equilibrium" (under certain assumptions about the learning model). Thus, in this class of games, the notion of stochastic stability offers a solution to the equilibrium selection problem.

Young's algorithm, however, is unwieldy and not in widespread use. Ellison (2000) suggests a simpler procedure for computing stochastically stable states. But both Young and Ellison fall short of presenting an algorithm to compute the stochastically stable distribution of a perturbed Markov process. The only known "algorithm" for computing a stochastically stable distribution, to our knowledge, is an iterative approach that falls out of the definition. However, since the computation of such distributions can be numerically unstable, such a sequence may appear to be converging to the *wrong* limit, or the limit may appear to be undefined (see Section 6). In this paper, we present the first exact algorithm for computing the stochastically stable distribution of a perturbed Markov process.

We use our algorithm to predict the long-run dynamics of simple learning models in sample repeated games. In particular, we replicate Young's adaptive learning model on several well-known two-player games. Using our algorithm, we can easily compute both the stochastically stable states and its corresponding distribution, enabling us to systematically investigate various choices of the parameters in this learning model.

This paper is organized as follows: Next, we introduce, by examples, the key constructions, which motivate the design of the algorithm. In Section 3, we review some basic definitions pertaining to Markov processes. We then formally define the quotient construction, which lies at the heart of our algorithm. In Section 5, we prove our main theorem, which yields an exact algorithm to compute the stochastically stable distribution of a perturbed Markov process. We conclude by applying this algorithm, assuming Young's adaptive learning model, to several well-known two-player games, investigating the equilibrium selection problem in these games.

## 2 The Algorithm at a High-Level

Our algorithm for computing the stochastically stable distribution of a perturbed Markov process is based on two fundamental operations on Markov processes, "scaling" transitions and "eliminating" states, which we combine in a "quotient" operation that effectively "collapses" states. Using such operations, we systematically simplify a perturbed Markov process, keeping track of the effects on the stochastically stable distribution, until we reach a trivial process (i.e., one with a single state). In this section, we describe these operations informally, illustrate them on simple examples, and present a high-level description of our algorithm.

**Note**: All our operations on a Markov process, specified by a Markov matrix $M$, are more naturally understood as operations on the displacement, $M - I$.

The "scaling" operation transforms $M$ via right-multiplication of $M-I$ by a suitable, invertible matrix $D$, called the "scaling" matrix, yielding $(M - I) D + I$. To illustrate, let $M_0 = \frac{1}{3} \begin{pmatrix} 0 & 1 & 1 \\ 2 & 1 & 2 \\ 1 & 1 & 0 \end{pmatrix}$. If we "scale" by $D = \begin{pmatrix} 1 & 0 & 0 \\ 0 & \frac{3}{2} & 0 \\ 0 & 0 & 1 \end{pmatrix}$, we obtain a new process given by $M_1 = \begin{pmatrix} 0 & \frac{1}{2} & \frac{1}{3} \\ \frac{2}{3} & 0 & \frac{2}{3} \\ \frac{1}{3} & \frac{1}{2} & 0 \end{pmatrix}$. The scaling operation effectively "speeds up" time in state 2 to eliminate the self-loop. This operation affects the stable distribution in a predictable manner. Specifically, $D$ maps the stable distribution of $M_1$, namely $\begin{pmatrix} \frac{3}{10} & \frac{2}{5} & \frac{3}{10} \end{pmatrix}^t$, to a multiple (i.e., $\frac{6}{5}$) of the stable distribution of $M_0$, namely $\begin{pmatrix} \frac{1}{4} & \frac{1}{2} & \frac{1}{4} \end{pmatrix}^t$. In Section 3.1, we introduce this notion formally as $D$-*equivalence*.

A more interesting example "scales" $M_1$ by $i = \begin{pmatrix} 1 & 0 & 0 \\ 0 & 1 & 0 \\ \frac{1}{3} & \frac{1}{2} & 1 \end{pmatrix}$ to obtain $M_2 = \frac{1}{9} \begin{pmatrix} 1 & 6 & 3 \\ 8 & 3 & 6 \\ 0 & 0 & 0 \end{pmatrix}$. This particular operation has the effect of eliminating all transitions into state 3. As above, $i$ maps the stable distribution of $M_2$, namely $\begin{pmatrix} \frac{3}{7} & \frac{4}{7} & 0 \end{pmatrix}^t$, to a multiple (i.e., $\frac{10}{7}$) of that of $M_1$. We can achieve a similar effect if we scale $M_1$ by $i^* = \begin{pmatrix} \frac{3}{4} & 0 & 0 \\ 0 & \frac{2}{3} & 0 \\ \frac{1}{4} & \frac{1}{3} & 1 \end{pmatrix}$ to obtain

$M_3 = \frac{1}{9} \begin{pmatrix} 3 & 4 & 3 \\ 6 & 5 & 6 \\ 0 & 0 & 0 \end{pmatrix}$. The matrix $M_3$ is simply $M_2$ "scaled" by $D = \begin{pmatrix} \frac{3}{4} & 0 & 0 \\ 0 & \frac{2}{3} & 0 \\ 0 & 0 & 1 \end{pmatrix}$. However, since the columns of $i^*$ sum to 1, $i^*$ now maps $\begin{pmatrix} \frac{2}{5} & \frac{3}{5} & 0 \end{pmatrix}^t$, the stable distribution of $M_3$, to that of $M_1$ exactly.

To summarize, "scaling" converts one Markov process into another that is "equivalent", in the sense that a simple transformation relates the corresponding stable distributions. We can generalize this operation to "eliminate" states. For example, $\bar{\imath} = \begin{pmatrix} 1 & 0 \\ 0 & 1 \\ 0 & 0 \end{pmatrix}$ maps the stable distribution, namely $\begin{pmatrix} \frac{2}{5} & \frac{3}{5} \end{pmatrix}^t$, of $M_4 = \frac{1}{9} \begin{pmatrix} 3 & 4 \\ 6 & 5 \end{pmatrix}$ to that of $M_3$. Thus, in a certain sense, $M_4$ can be considered "equivalent" to $M_3$ via $\bar{\imath}$. In addition, if we let $p = \begin{pmatrix} 1 & 0 & \frac{1}{3} \\ 0 & 1 & \frac{2}{3} \end{pmatrix}$, then $M_4 = p(M_3 - I)\bar{\imath} + I$. This transformation, which is an instance of the quotient operation defined in Section 4, is a generalization of "scaling," where the scaling matrix need only have full column rank.

Putting these two types of operations together— "scaling" transitions into states to 0 (as in the computations of $M_2$ and $M_3$) and "eliminating" states (as in the computation of $M_4$)—we can effectively "collapse" several states into a single state. For example, consider $M = \begin{pmatrix} \frac{1}{4} & \frac{1}{4} & \frac{1}{2} \\ 0 & \frac{1}{4} & 0 \\ \frac{3}{4} & \frac{1}{2} & \frac{1}{2} \end{pmatrix}$. If we let $\tilde{\imath} = \begin{pmatrix} 1 & 0 \\ 0 & 1 \\ \frac{3}{2} & 1 \end{pmatrix}$ and $\tilde{p} = \begin{pmatrix} 1 & 0 & 1 \\ 0 & 1 & 0 \end{pmatrix}$, then $\widetilde{M} = \tilde{p}(M - I)\tilde{\imath} + I = \begin{pmatrix} 1 & \frac{3}{4} \\ 0 & \frac{1}{4} \end{pmatrix}$. Via this quotient operation, states 1 and 3 in $M$ are combined into a single state in $\widetilde{M}$. More precisely, state 3 is collapsed into state 1. Moreover, $\tilde{\imath}$ maps the stable distribution of $\widetilde{M}$ to a multiple of that of $M$. And, as above, by first normalizing the columns of $\tilde{\imath}$ to obtain $\tilde{\imath}^* = \begin{pmatrix} \frac{2}{5} & 0 \\ 0 & \frac{1}{2} \\ \frac{3}{5} & \frac{1}{2} \end{pmatrix}$, the stable distribution of $\tilde{p}(M - I)\tilde{\imath}^* + I = \begin{pmatrix} 1 & \frac{3}{8} \\ 0 & \frac{5}{8} \end{pmatrix}$ is mapped exactly to that of $M$ by $\tilde{\imath}^*$. This latter operation is an example of a *normalized quotient*, defined in Section 4.

We now briefly describe our algorithm to compute the stochastically stable distribution of a perturbed Markov process. First, the perturbation rate is set to 0 to obtain an unperturbed process. Next, this unperturbed process is examined to determine an appropriate set of states to collapse. The quotient operators $p$ and $i^*$ from the *unperturbed* process are then applied to the *perturbed* process to obtain an equivalent perturbed Markov process (i.e., one whose stochastically stable distribution is mapped to that of the original by $i^*$) with strictly fewer states. Scaling is performed as necessary to ensure that the resulting process is not degenerate. This procedure is iterated, accumulating the $i^*$'s along the way, until the process is trivial. The composite of the $i^*$'s yields the stochastically stable distribution.

## 3 Markov Matrices

Let $J = [1, \ldots, 1]$ denote a row vector of 1s. An $n \times n$ matrix is called *Markov* iff $M \geq 0$ and $JM = J$: i.e., all columns sum to 1. A *distribution* is a vector $v \geq 0$ s.t. $Jv = 1$. A Markov process $\mathcal{M}$ is uniquely characterized by a Markov matrix and an initial distribution. Throughout, we assume a discrete-time, finite-state Markov process. Algebraically, each state corresponds to a row (and column) index in $M$; geometrically, each state corresponds to a vertex of the standard simplex.

Observe that a Markov matrix $M$ may be viewed as a weighted, directed graph whose weights are in $[0, 1]$ such that the sum of the outgoing edges from any node is 1. Under this interpretation, we can partition the vertices into *communicating classes*, which correspond to its strongly connected components (SCC). Alternatively, a communicating class is a maximal set of states among which there is positive probability of eventually reaching any state from any other state. A subset of states is said to be *invariant* if the probability of ever transitioning away from the set is 0. A *closed* class is a SCC out of which there are no outgoing edges to any other SCC; that is, an invariant, communicating class. States that do not belong to any closed class are called *transient*, since there is positive probability of transitioning away from such states without ever returning. A Markov chain is said to be *reducible* if it contains more than one communicating class; otherwise it is *irreducible*. We call a Markov chain *regular* if it contains exactly one closed class.

Given a Markov matrix $M$, a *stable* distribution of $M$ is one which is also an eigenvector with eigenvalue 1: i.e., $Mv = v$. If $\dim M = n$ and $\Delta$ is the standard $n$–simplex, then the set of stable distributions of $M$, $\text{stab}(M) = \ker(M - I) \cap \Delta$. It is well-known that if $M > 0$, then $|\text{stab}(M)| = 1$ and the associated Markov process converges to the unique stable distribution, regardless of the initial distribution. In Section 4, we show that $|\text{stab}(M)| = 1$ even if $M$ is only regular.

## 3.1 Equivalence

We say that two matrices $M_1$ and $M_2$ are *equivalent* iff:
$$\ker(M_1 - I) = \ker(M_2 - I) \quad (1)$$

More generally, we say that $M_2$ is *D-equivalent* to $M_1$ iff there exists an injective mapping $D$ such that:
$$\ker(M_1 - I) = D\ker(M_2 - I) \quad (2)$$

This equivalence condition says that $D$ maps $\ker(M_2 - I)$ onto $\ker(M_1 - I)$, implying that $D$ is in fact a bijective mapping between the two kernels. Note that D-equivalence is *not* an equivalence relation. It is a partial order, since it is not symmetric, but it is reflexive (choose $D = I$) and transitive (if $M_2$ is D-equivalent to $M_1$ and $M_3$ is $D'$-equivalent to $M_2$, then $M_3$ is $DD'$-equivalent to $M_1$). In this section, we give sufficient conditions for two Markov matrices to be D-equivalent. To do so, we rely on the following preliminary lemma.

**Lemma 3.1** *Given three $n \times n$ matrices $A$, $B$, and $C$, such that $AB = C$, with $B$ invertible, $\ker(A) = B\ker(C)$.*

**Proof 3.1** If $v \in \ker(A)$, then $B^{-1}v \in \ker(C)$, so that $B^{-1}\ker(A) \subseteq \ker(C)$; equivalently, $\ker(A) \subseteq B\ker(C)$. If $v \in \ker(C)$, then $Bv \in \ker(A)$, so that $B\ker(C) \subseteq \ker(A)$. □

Given a Markov matrix $M$ and diagonal matrix $D$ with $0 < D_{ii} \leq (1 - M_{ii})^{-1}$, we define $M_D \equiv (M - I)D + I = MD + (1 - D)I$.

**Lemma 3.2** *Given a Markov matrix $M$, $M_D$ is a Markov matrix that is D-equivalent to $M$.*

**Proof 3.2** The fact that $M_D$ is D-equivalent to $M$ follows directly from Lemma 3.1, noting that $D$ is invertible (choose $A = M - I$, $B = D$, and $C = (M - I)D$). It remains to show that $M_D$ is Markov. First, $JM_D = J(M - I)D + J = (J - J)D + J = J$: i.e., the columns of $M_D$ sum to 1. Second, all the off-diagonal entries of $M_D$ are nonnegative, since, for $i \neq j$, $(M_D)_{ij} = (MD)_{ij} - D_{ij} + I_{ij} = (MD)_{ij}$ and $M, D \geq 0$. Third, all the diagonal entries of $M_D$ are nonnegative, since $(M_D)_{ii} = (MD)_{ii} - D_{ii} + 1 = M_{ii}D_{ii} - D_{ii} + 1 = (M_{ii} - 1)D_{ii} + 1 = 1 - (1 - M_{ii})D_{ii}$, so that $0 \leq (M_D)_{ii}$ iff $(1 - M_{ii})D_{ii} \leq 1$ iff $D_{ii} \leq (1 - M_{ii})^{-1}$, which holds by assumption. □

We rely on the following corollary in our analysis of perturbed Markov processes.

**Corollary 3.3** *Given a Markov matrix $M$, if $0 < \epsilon \leq \min_i (1 - M_{ii})^{-1}$, then $M_\epsilon \equiv (M - I)\epsilon + I$ is a parameterized family of Markov matrices equivalent to $M$.*

**Proof 3.3** Letting $D_{ii} = \epsilon$, for all $1 \leq i \leq n$, $D$ is simply scalar multiplication. By Lemma 3.2, $\ker(M - I)$ is a scalar multiple of $\ker(M_\epsilon - I)$. But then these kernels are equal, since they are both subspaces of $\mathbb{R}^n$. □

Intuitively, for $\epsilon < 1$, $M_\epsilon$ represents the associated Markov process where time has been "slowed down"; at each step of the new process, only with probability $\epsilon$ is a decision made to transition according to the original process $M$. More generally, in $M_D$, time can be scaled differently in each dimension. In particular, by taking $D_{ii} = (1 - M_{ii})^{-1} > 1$ (its maximum value), time is "sped up" to eliminate self–loops at state $i$, for all $1 \leq i \leq n$ where $M_{ii} \neq 1$.

## 4 The Quotient Construction

In this section, we define the quotient construction which lies at the heart of our algorithm for computing stochastically stable distributions. This construction formalizes and generalizes common "folk wisdom" regarding the collapsing of communicating classes in Markov processes.

Before proceeding, we state without proof the following fact from linear algebra.

**Lemma 4.1** *Consider the submatrix $\overline{M}$ of a Markov matrix $M$ on a set of states $\mathcal{P}$. If $\mathcal{P}$ does not contain a closed class of $M$, then $I - \overline{M}$ is invertible, $\overline{M}^k \to 0$, and $(I - \overline{M})^{-1} = \lim_{i \to \infty} \sum_{j=0}^{i-1} \overline{M}^j \equiv \sum_{j=0}^{\infty} \overline{M}^j$.*

Assume $M = \begin{pmatrix} \widetilde{M} & \overline{N} \\ \widetilde{N} & \overline{M} \end{pmatrix}$ is partitioned via the states in $\mathcal{P}$ so that $\widetilde{M}$ describes the transitions among the states in $\overline{\mathcal{P}}$; $\overline{M}$ describes the transitions among the states in $\mathcal{P}$; $\overline{N}$ describes the transitions out of the states in $\mathcal{P}$ into the states in $\overline{\mathcal{P}}$; and $\widetilde{N}$ describes the transitions out of the states in $\overline{\mathcal{P}}$ into the states in $\mathcal{P}$. Our quotient construction is based on column elimination. Given any set of states $\mathcal{P}$ that does not contain an entire closed class of $M$, we can eliminate the transitions from $\overline{\mathcal{P}}$ into $\mathcal{P}$, while changing the set of stable distributions in a predictable way. Specifically, by column elimination on $M - I$, we obtain a new matrix $M' = \begin{pmatrix} \widehat{M} & \overline{N} \\ 0 & \overline{M} \end{pmatrix}$ such that $\text{stab}(M)$ is in 1-to-1 correspondence with $\text{stab}(\widehat{M})$ (i.e., we exhibit a bijective mapping from $\text{stab}(M)$ to $\text{stab}(\widehat{M})$).

### 4.1 Quotients

Column elimination on $M - I$ is equivalent to a factorization of the form

$$(M - I) \begin{pmatrix} I & 0 \\ A & I \end{pmatrix} = \begin{pmatrix} \widetilde{M} - I & \overline{N} \\ \widetilde{N} & \overline{M} - I \end{pmatrix} \begin{pmatrix} I & 0 \\ A & I \end{pmatrix} = \begin{pmatrix} B & C \\ 0 & D \end{pmatrix} \quad (3)$$

By Lemma 4.1, since $\mathcal{P}$ does not contain an entire closed class of $M$, $I - \overline{M}$ is invertible, so that $C = \overline{N}$, $D = \overline{M} - I$, $A = (I - \overline{M})^{-1} \widetilde{N}$, and $B = \widetilde{M} + \overline{N} (I - \overline{M})^{-1} \widetilde{N} - I = \widehat{M} - I$. We call $\widehat{M}$ the *quotient* of $M$ with respect to $\mathcal{P}$. In particular:

**Lemma 4.2** *The quotient $\widehat{M} = \widetilde{M} + \overline{N}(I - \overline{M})^{-1} \widetilde{N}$.*

Intuitively, we may view this construction as forcing transitions among the states in $\mathcal{P}$ to occur "instantaneously." Observe that $\overline{N}(I - \overline{M})^{-1} \widetilde{N} = \overline{N} \left( \sum_{j=0}^{\infty} \overline{M}^j \right) \widetilde{N}$ represents the probability of going from one state in $\overline{\mathcal{P}}$ to another, visiting the states in $\mathcal{P}$ an arbitrary number of times. Similarly, $\widetilde{M}$ represents the probability of traversing only a single edge from a state in $\overline{\mathcal{P}}$ to another. Thus, $\widehat{M}_{ij}$ represents the probability of going from state $i$ to state $j$, both in $\overline{\mathcal{P}}$, via a walk whose interior states are only in $\mathcal{P}$, including the trivial case of traversing only the edge from $i$ to $j$.

More importantly, we can give an alternative characterization of this quotient construction. Let $k$ be the number of states in $\overline{\mathcal{P}}$. Define the following: a $k \times n$-dimensional matrix

$$p = \begin{pmatrix} I & \overline{N}(I - \overline{M})^{-1} \end{pmatrix} \quad (4)$$

and an $n \times k$-dimensional matrix

$$i = \begin{pmatrix} I \\ (I - \overline{M})^{-1} \widetilde{N} \end{pmatrix} \quad (5)$$

The matrix $p$ is a linear, projection operator (i.e., a surjective mapping), since it has full row rank. The matrix $i$ is an linear, inclusion operation (i.e., an injective mapping), since it has full column rank. **Note**: The formulas given for $p$ and $i$ are written up to a permutation of the columns of $p$ and the rows of $i$, which depends on the choice of $\mathcal{P}$.

Now

$$(M - I) i = \begin{pmatrix} \widetilde{M} - I & \overline{N} \\ \widetilde{N} & \overline{M} - I \end{pmatrix} \begin{pmatrix} I \\ (I - \overline{M})^{-1} \widetilde{N} \end{pmatrix} = \begin{pmatrix} \widehat{M} - I \\ 0 \end{pmatrix} \quad (6)$$

so that

$$\begin{aligned} p(M-I)i + I &= \begin{pmatrix} I & \overline{N}(I - \overline{M})^{-1} \end{pmatrix} \begin{pmatrix} \widehat{M} - I \\ 0 \end{pmatrix} \quad (7) \\ &= (\widehat{M} - I) + I = \widehat{M} \quad (8) \end{aligned}$$

In other words, up to translations by the identity, $\widehat{M}$ can be computed by composition with $p$ and $i$.

Note that the matrices $p$ and $i$ are both positive. In addition, $p$ is a mapping between distributions, since its columns sum to 1: $J\overline{M} + J\overline{N} = J$ (since $JM = J$), so $J\overline{N} = J - J\overline{M} = JI - J\overline{M} = J(I - \overline{M})$, $J\overline{N}(I - \overline{M})^{-1} = J$, and $Jp = J$.

**Lemma 4.3** *If $M$ is Markov, then $\widehat{M}$ is Markov.*

**Proof 4.3** First, since the columns of $p$ sum to 1, the columns of $\widehat{M}$ sum to 1: $Jp(M - I) = J(M - I) = (JM - JI) = (J - J) = 0$, so $J\widehat{M} = J[p(M - I)i + I] = Jp(M - I)i + J = 0i + J = J$. Second, Lemma 4.2 implies that $\widehat{M} \geq 0$. Therefore, $\widehat{M}$ is Markov. □

As above, let $k$ be the number of states in $\overline{\mathcal{P}}$, and assume that $\mathcal{P} = \{k+1, \ldots, n\}$. Let $I$ be $k$-dimensional, and define the $k \times n$ dimensional matrix $\pi = \begin{pmatrix} I & 0 \end{pmatrix}$. We say that $w \in \mathbb{R}^n$ is an *extension* of a vector $v \in \mathbb{R}^k$ iff $v = \pi w$. **Note**: The formula given for $\pi$ is written up to a permutation of the columns of $\pi$, which depends on the choice of $\mathcal{P}$.

**Theorem 4.4** *Given a Markov matrix $M$, $\widehat{M}$ is i-equivalent to $M$. In particular, the eigenvector $iv \in \ker(M - I)$ is the unique extension of an eigenvector $v \in \ker(\widehat{M} - I)$.*

**Proof 4.4** By Equation 3,

$$(M - I) \begin{pmatrix} I & 0 \\ (I - \overline{M})^{-1} \widetilde{N} & I \end{pmatrix} = \begin{pmatrix} \widehat{M} - I & \overline{N} \\ 0 & \overline{M} - I \end{pmatrix} \quad (9)$$

Therefore,

$$\begin{aligned} \ker(M - I) &= \begin{pmatrix} I & 0 \\ (I - \overline{M})^{-1} \widetilde{N} & I \end{pmatrix} \ker \begin{pmatrix} \widehat{M} - I & \overline{N} \\ 0 & \overline{M} - I \end{pmatrix} \quad (10) \\ &= \begin{pmatrix} I & 0 \\ (I - \overline{M})^{-1} \widetilde{N} & I \end{pmatrix} \left\{ \begin{pmatrix} v \\ 0 \end{pmatrix} \mid v \in \ker(\widehat{M} - I) \right\} \quad (11) \\ &= \begin{pmatrix} I \\ (I - \overline{M})^{-1} \widetilde{N} \end{pmatrix} \ker(\widehat{M} - I). \quad (12) \end{aligned}$$

Equation 10 follows from Lemma 3.1. Equation 11 follows via the following reasoning. If $\begin{pmatrix} v \\ w \end{pmatrix} \in \ker \begin{pmatrix} \widehat{M} - I & \overline{N} \\ 0 & \overline{M} - I \end{pmatrix}$, then $(\widehat{M} - I)v + \overline{N}w = 0$ and $(\overline{M} - I)w = 0$. Since $\overline{M} - I$ is invertible, $w = 0$. But then $(\widehat{M} - I)v = 0$: i.e., $v \in \ker(\widehat{M} - I)$. Conversely, if $v \in \ker(\widehat{M} - I)$, then $\begin{pmatrix} v \\ 0 \end{pmatrix} \in \ker \begin{pmatrix} \widehat{M} - I & \overline{N} \\ 0 & \overline{M} - I \end{pmatrix}$.

Since $\pi i = I$, it follows that $\pi i v = v$ so that $iv$ is an extension of $v \in \ker(\widehat{M} - I)$. Suppose that $w \in$

$\ker (M - I)$ is another extension of $v \in \ker \left(\widehat{M} - I\right)$: that is, $v = \pi w$. Since $i$ maps $\ker \left(\widehat{M} - I\right)$ onto $\ker (M - I)$, there exists $v' \in \ker \left(\widehat{M} - I\right)$ such that $w = iv'$. Now $v = \pi w = \pi i v' = v'$. Therefore, $w = iv$ so that $iv \in \ker (M - I)$ is the unique extension of $v \in \ker \left(\widehat{M} - I\right)$. □

**Theorem 4.5** *If $M$ is a regular Markov matrix, then $\widehat{M}$ is also a regular Markov matrix.*

**Proof 4.5** By Lemma 4.3, $\widehat{M}$ is Markov. Since the quotient construction preserves all path connections among the states in $\overline{\mathcal{P}}$, and since regularity is defined in terms of path connectivity, the result follows. □

#### 4.1.1 Normalized Quotients

If $i$ is the inclusion operator corresponding to the quotient $\widehat{M}$ with respect to some set of states $\mathcal{P}$, let $d$ be the diagonal matrix such that $Jd = Ji$: that is, the diagonal entries of $d$ correspond to the column sums of $i$. We call $\widehat{M}^* \equiv \widehat{M}_{d^{-1}}$ the *normalized quotient* of $M$ with respect to $\mathcal{P}$, $d^{-1}$ the *normalizing matrix* of the quotient, and $i^* = id^{-1}$ the *normalized inclusion operator*. Note that $i^*$ is injective, since $i$ is injective. Moreover, $Ji^* = Jid^{-1} = Jdd^{-1} = J$, that is, the columns of $i^*$, which are precisely the columns of $i$ normalized, sum to 1.

**Corollary 4.6** *Given a Markov matrix $M$, $\widehat{M}^*$ is $i^*$-equivalent to $M$. In particular, the stable distributions of $\widehat{M}^*$ are mapped bijectively to those of $M$ via $i^*$.*

**Proof 4.6** Choosing $M = \widehat{M}$ and $D = d^{-1}$ in Lemma 3.2, $\widehat{M}_{d^{-1}}$ is $d^{-1}$-equivalent to $\widehat{M}$. By Theorem 4.4, $\widehat{M}$ is $i$-equivalent to $M$. Hence, by transitivity of the relation, $\widehat{M}^* = \widehat{M}_{d^{-1}}$ is $i^* = id^{-1}$-equivalent to $M$: that is, $i^*$ maps $\ker \left(\widehat{M}^* - I\right)$ onto $\ker (M - I)$. Now, since $i^*$ is injective, it is in fact a bijective mapping between $\ker \left(\widehat{M}^* - I\right)$ and $\ker(M - I)$. But $i^*$ is positive with columns that sum to 1. Therefore, $i^*$ restricts to a bijection from $\text{stab}\left(\widehat{M}^*\right)$ to $\text{stab}(M)$. □

#### 4.1.2 Maximal Quotients

Given a Markov matrix $M$, let $\mathcal{P}$ contain all states but one member of each closed class. In particular, $k = |\overline{\mathcal{P}}|$ is the number of distinct closed classes. A *maximal quotient* is the result of applying the quotient construction to eliminate any states in such a $\mathcal{P}$.

**Theorem 4.7** *Given a Markov matrix $M$, any maximal quotient $\widehat{M} = I$. Notably, $\ker \left(\widehat{M} - I\right) = \mathbb{R}^k$.*

**Proof 4.7** Since there are no paths among closed classes, any maximal quotient $\widehat{M} = I$. (Here, $I$, and hence $\widehat{M}$, are $k$-dimensional.) It follows immediately that $\ker \left(\widehat{M} - I\right) = \mathbb{R}^k$. □

**Corollary 4.8** *Every regular Markov matrix $M$ has a unique stable distribution $v$.*

**Proof 4.8** Since $M$ is regular, $k = 1$. By Theorem 4.7, $\dim \ker \left(\widehat{M} - I\right) = 1$, for any maximal quotient $\widehat{M}$. By Theorem 4.4, $\dim \ker (M - I) = 1$. In particular, $|\text{stab}(M)| = 1$. □

While the following corollary is well-known (e.g., Horn and Johnson (1985)), the quotient construction provides a straightforward alternate proof of this result.

**Corollary 4.9** *Every irreducible Markov matrix $M$ has a unique stable distribution $v > 0$.*

**Proof 4.9** Since every irreducible Markov matrix is regular, by Corollary 4.8, $M$ has a unique stable distribution. It remains to show that $v > 0$. We proceed by contradiction. Assume, without loss of generality, that $v = \begin{pmatrix} 0 \\ \overline{v} \end{pmatrix}$, with $\overline{v} > 0$, and $M = \begin{pmatrix} A & B \\ C & D \end{pmatrix}$, so that $\begin{pmatrix} A & B \\ C & D \end{pmatrix} \begin{pmatrix} 0 \\ \overline{v} \end{pmatrix} = \begin{pmatrix} 0 \\ \overline{v} \end{pmatrix}$. Since $B \geq 0$, while $B\overline{v} = 0$ and $\overline{v} > 0$, it must be the case that $B = 0$, contradicting the irreducibility of $M$. □

### 4.2 Naturality of the Quotient Construction

Given a Markov matrix $M$, let $\mathcal{P}$ be any set of states that does not contain an entire closed class and compute the associated quotient operators $p$ and $i$. We denote the quotient of $M$ with respect to $\mathcal{P}$ as the triple $\left(\widehat{M}, p, i\right)$. This construction is natural, in the sense of category theory:

**Theorem 4.10** *If $\mathcal{P} = \mathcal{P}_1 \cup \mathcal{P}_2$; $(M_1, p_1, i_1)$ is the quotient of $M$ with respect to $\mathcal{P}_1$; $(M_2, p_2, i_2)$ is the quotient of $M_1$ with respect to $\mathcal{P}_2$; $\left(\widehat{M}, p, i\right)$ is the quotient of $M$ with respect to $\mathcal{P}$; then $p = p_2 p_1$ and $i = i_1 i_2$.*

**Proof 4.10** Let $T(\pi, \mathcal{P}) = \min \{i \mid \pi_i \in \overline{\mathcal{P}}\}$ be the first time the trajectory $\pi$ visits states outside $\mathcal{P}$ (since $\mathcal{P}$ does not contain a closed class, this time is well-defined). Let $S(\pi, \mathcal{P}) = \pi_{T(\pi, \mathcal{P})}$ be the first such state that is visited. Since $p_{ij}$ represents the probability that $i$ is the first state in $\overline{\mathcal{P}}$ that is visited along a trajectory $\pi$ that originates at $j$, we can write $p_{ij} = \Pr \{S(\pi, \mathcal{P}) = i \mid \pi_0 = j\}$. Now $\overline{\mathcal{P}} \subseteq \overline{\mathcal{P}}_1$; hence, any trajectory from $j$ to $\overline{\mathcal{P}}$ must enter $\overline{\mathcal{P}}_1$ at some state

$k$. There is exactly one such first state, so that $p_{ij} = \sum_k \Pr\{S(\pi, \mathcal{P}_1 \cup \mathcal{P}_2) = i, S(\pi, \mathcal{P}_1) = k \mid \pi_0 = j\} = \sum_k (p_2)_{ik}(p_1)_{kj}$. Therefore, $p = p_2 p_1$.

Finally, by Theorem 4.4, $i(v)$ is the unique extension of an eigenvector $v \in \ker\left(\widehat{M} - I\right)$ to an eigenvector in $\ker(M - I)$. Likewise, $i_1 i_2(v)$ is an extension of an eigenvector $v \in \ker(M_2 - I)$ to an eigenvector in $\ker(M_1 - I)$ to an eigenvector in $\ker(M - I)$. Therefore, since $\widehat{\widehat{M}} = M_2$, by uniqueness, $i_1 i_2(v) = i(v)$. □

## 5 Perturbed Markov Processes

For sufficiently small $\epsilon \geq 0$, a *perturbed Markov process* $M_\epsilon$ is a family of Markov processes whose entries converge exponentially fast. That is, either $(M_\epsilon)_{ij} = 0$ or $(M_\epsilon)_{ij} = \epsilon^{r_{ij}} c_{ij}(\epsilon)$, for $r_{ij}$ integers and $c_{ij}(\epsilon)$ continuous, strictly positive functions of $\epsilon$. By convention, if $(M_\epsilon)_{ij} = 0$, we let $r_{ij} = \infty$ and $\epsilon^{r_{ij}} = 0$. Note that, if $M_\epsilon$ is a sufficiently smooth function of $\epsilon$, then $r_{ij} = \min\left\{r \mid \left((M_\epsilon)_{ij}\right)^{(r)}(0) \neq 0\right\}$. The value $r_{ij}$ is called the *degree* or *resistance* of edge $(i, j)$.

If $M_\epsilon$ is regular for all $\epsilon > 0$, this perturbed Markov process is called *regular*.[1] Let $v_\epsilon$ denote the unique stable distribution of $M_\epsilon$. Such a distribution necessarily exists and is unique by Corollary 4.8.

**Lemma 5.1** *If $M_\epsilon$ is a regular perturbed Markov process, then the entries of $v_\epsilon$ converge exponentially fast.*

**Proof 5.1** Omitted. □

By Lemma 5.1, we can define the *stochastically stable distribution* of $M_\epsilon$ as $v_0 \equiv \lim_{\epsilon \to 0} v_\epsilon$ and the *stochastically stable states* of $M_\epsilon$ as $\mathrm{supp}_{v_0}$. We are interested in calculating this stochastically stable distribution. We say that two regular perturbed Markov processes are *equivalent* if they have the same stochastically stable distribution and path connections (i.e., there is a path between two given states in one process iff there is a path between the same two states in the other). Note that, without any increase in generality, we can allow $r_{ij}$ to be rational. Given a regular perturbed Markov process, by reparameterization, we can obtain an equivalent (regular) perturbed Markov process with integer exponents.

**Lemma 5.2** *The stochastically stable distribution of a regular perturbed Markov process $M_\epsilon$ depends only on the leading coefficients $c_{ij}(0)$ of the functions $c_{ij}(\epsilon)$. In particular, $M_\epsilon$ is equivalent to one in which all the off-diagonal functions $c_{ij}(\epsilon)$ are constant, and the diagonal entries have 0-resistance.*

**Proof 5.2** Omitted. □

**Lemma 5.3** *A regular perturbed Markov process is equivalent to one of the form $M_\epsilon = M_0 + \epsilon N_\epsilon$, for $M_0 = I + N$ with $N \neq 0$ and $N_\epsilon$ polynomial in $\epsilon$.*

**Proof 5.3** Omitted. □

The proofs of Lemmas 5.1 and 5.2 are technical arguments that rely on the Markov Chain Tree Theorem (see, for example, Friedlin and Wentzell (1984)), while Lemma 5.3 is a relatively straightforward application of Corollary 3.3. We now state a penultimate technical Lemma, which also exploits the Markov Chain Tree Theorem.

**Lemma 5.4** *Let $M_\epsilon = M_0 + N_\epsilon \epsilon$ be a regular perturbed Markov process, as in Lemma 5.3. Choose one non-trivial communicating class of $M_0$ (i.e., containing more than one element), let $\mathcal{P}_0$ be all but one representative from that set, and let $\left(\widehat{M_\epsilon}, p_\epsilon, i_\epsilon\right)$ be the quotient with respect to $\mathcal{P}_0$ of $M_\epsilon$. Then $\widehat{M_\epsilon}$ is equivalent to $\widetilde{M_\epsilon} \equiv p_0(M_\epsilon - I)i_0 + I$.*

**Proof 5.4** Omitted. □

Our algorithm for computing the stochastically stable distribution of a perturbed Markov process requires one final, easily proven Lemma.

**Lemma 5.5** *Let $M_\epsilon$ be a regular perturbed Markov process for which each communicating class of $M_0$ is trivial. If $\mathcal{T}$ is the set of transient states with indicator function $\chi_\mathcal{T}$, $i_\epsilon$ is a diagonal matrix with $(i_\epsilon)_{j,j} = \epsilon^{\chi_\mathcal{T}(j)}$, and $D = \frac{i_\epsilon}{\epsilon}$, then $(M_\epsilon)_D$ is $i_\epsilon$-equivalent to $M_\epsilon$ with $i_0$ mapping the stochastically stable distribution of $(M_\epsilon)_D$ to that of $M_\epsilon$.*

**Proof 5.5** Notice that $(M_\epsilon)_D = \left((M_\epsilon)_{\frac{1}{\epsilon}}\right)_{i_\epsilon}$. Then by Lemma 3.2, for $\epsilon > 0$, $(M_\epsilon)_D$ is $i_\epsilon$-equivalent to $(M_\epsilon)_{\frac{1}{\epsilon}}$, which, by Corollary 3.3, is equivalent to $M_\epsilon$. □

We are now ready to prove the main result of this paper, which forms the heart of our algorithm.

**Theorem 5.6** *Let $M_\epsilon = M_0 + N_\epsilon \epsilon$ be a regular perturbed Markov process, as in Lemma 5.3. Take $\mathcal{P}_0$ to contain all states but one representative of each communicating class in $M_0$, and let $\left(\widehat{M_0^*}, p_0, i_0^*\right)$ be the normalized quotient with respect to $\mathcal{P}_0$ of $M_0$. Then* $\widehat{\widehat{M}}_\epsilon^* \equiv p_0 (M_\epsilon - I) i_0^* + I$ *is a regular perturbed Markov process and $i_0^*$ maps the stochastically stable distribution of $\widehat{\widehat{M}}_\epsilon^*$ to that of $M_\epsilon$.*

---

[1] This generalizes the usual definition of a perturbed Markov process, which requires that $M_\epsilon$ be irreducible—i.e., there exists one closed class of which all states are elements—rather than merely regular—i.e., there exists one closed class, but some states may be transient.

**Proof 5.6** Let $\left(\widehat{M}_\epsilon, p_\epsilon, i_\epsilon\right)$ be the quotient with respect to $\mathcal{P}_0$ of $M_\epsilon$ with normalizing matrix $D_\epsilon$. Notice that this implies that $\left(\widehat{M}_0, p_0, i_0\right)$ is the quotient with respect to $\mathcal{P}_0$ of $M_0$ with normalizing matrix $D_0$.

Now, assume that $M_0$ has $k$ non-trivial communicating classes, and $\left(\widehat{M}_{j,\epsilon}, p_{j,\epsilon}, i_{j,\epsilon}\right)$ is the result of eliminating all but the chosen member of the $j$-th (non-trivial) communicating class from $\widehat{M}_{j-1,\epsilon}$, where $\widehat{M}_{0,\epsilon} \equiv M_\epsilon$. By Theorem 4.10, $\widehat{M}_{k,\epsilon} = \widehat{\widehat{M}}_\epsilon$ and $\widehat{M}_{j,\epsilon} = p_{j,\epsilon} \ldots p_{1,\epsilon} (M_\epsilon - I) i_{1,\epsilon} \ldots i_{j,\epsilon} + I$. Applying Lemma 5.4 $k$ times gives that $\widehat{\widehat{M}}_\epsilon$ is equivalent to $p_{k,0} p_{k-1,\epsilon} \ldots p_{1,\epsilon} (M_\epsilon - I) i_{1,\epsilon} \ldots i_{k-1,\epsilon} i_{k,0} + I$, etc., which is equivalent to $p_{k,0} \ldots p_{1,0} (M_\epsilon - I) i_{1,0} \ldots i_{k,0} + I$, which, by Theorem 4.10 again, is $\widehat{\widehat{M}}_\epsilon \equiv p_0 (M_\epsilon - I) i_0 + I$.

Now consider the associated normalized quotient, $\left(\widehat{M}_\epsilon^*, p_\epsilon, i_\epsilon^*\right)$, so that $i_\epsilon^* = i_\epsilon D_\epsilon$. We claim that $\widehat{M}_\epsilon^*$ is equivalent to $\widehat{\widehat{M}}_\epsilon^*$ as well. Let $\widehat{v}_\epsilon$ and $\widehat{v}_\epsilon^*$ be the stable distributions of $\left(\widehat{M}_\epsilon, p_\epsilon, i_\epsilon\right)$ and $\left(\widehat{M}_\epsilon^*, p_\epsilon, i_\epsilon^*\right)$, respectively, while $\widehat{\widehat{v}}_\epsilon$ and $\widehat{\widehat{v}}_\epsilon^*$ are the stable distributions of $\left(\widehat{\widehat{M}}_\epsilon, p_\epsilon, i_\epsilon\right)$ and $\left(\widehat{\widehat{M}}_\epsilon^*, p_\epsilon, i_\epsilon^*\right)$. Since all these processes are regular, we know that $\widehat{v}_\epsilon = D_\epsilon \widehat{v}_\epsilon^* / |D_\epsilon \widehat{v}_\epsilon^*|_1$ and $\widehat{\widehat{v}}_\epsilon = D_0 \widehat{\widehat{v}}_\epsilon^* / \left|D_0 \widehat{\widehat{v}}_\epsilon^*\right|_1$, for $\epsilon > 0$. In particular, taking limits, $\widehat{v}_0 = D_0 \widehat{v}_0^* / |D_0 \widehat{v}_0^*|_1$ and $\widehat{\widehat{v}}_0 = D_0 \widehat{\widehat{v}}_0^* / \left|D_0 \widehat{\widehat{v}}_0^*\right|_1$. Since $\widehat{M}_\epsilon$ is equivalent to $\widehat{\widehat{M}}_\epsilon$, $\widehat{v}_0 = \widehat{\widehat{v}}_0$. Thus, $D_0 \widehat{\widehat{v}}_0^* / \left|D_0 \widehat{\widehat{v}}_0^*\right|_1 = D_0 \widehat{v}_0^* / |D_0 \widehat{v}_0^*|_1$. Since the column sums of $i_\epsilon$ are positive for $\epsilon \geq 0$, $D_0$ is invertible and $\widehat{v}_0 \propto \widehat{\widehat{v}}_0$. Since they both are distributions, $\widehat{v}_0 = \widehat{\widehat{v}}_0$, so that $\widehat{M}_\epsilon^*$ is equivalent to $\widehat{\widehat{M}}_\epsilon^*$.

Now compare $v_\epsilon$ and $\widehat{v}_\epsilon^*$, the unique stable distributions of $M_\epsilon$ and $\widehat{M}_\epsilon^*$, respectively. By Corollary 4.6, $v_\epsilon = i_\epsilon^* \widehat{v}_\epsilon$, so that by taking limits $v_0 = i_0^* \widehat{v}_0 = i_0^* \widehat{\widehat{v}}_0$, as required.

Since the entries of $\widehat{\widehat{M}}_\epsilon^*$ are linear combinations of the entries of $M_\epsilon$, they converge exponentially. Since $M_\epsilon$ is regular, so is $\widehat{M}_\epsilon$, and $\widehat{M}_\epsilon^*$. Since $\widehat{\widehat{M}}_\epsilon$ has the same paths as $\widehat{M}_\epsilon^*$, it is also regular. □

Theorem 5.6, together with Lemmas 5.3 and 5.5, yield an exact algorithm to compute the stochastically sta-

**Algorithm 1** An Algorithm to Compute the Stochastically Stable Distribution of a Pertubed Markov Process.
1: $s = $ IdentityMatrix($M_\epsilon$.getDimension())
2: $M_\epsilon$.dropHighOrderTerms()
3: $M_\epsilon$.scale()
4: **while** $M_\epsilon$.numClosedClasses() > 1 **do**
5:    $M_0 = M_\epsilon$.getConstantTerms()
6:    $Q = $ getStatesExceptCommClassReps($M_0$)
7:    **if** $Q$.isEmpty() **then**
8:      $(T, N) = $ partitionStates($M_0$) {*partition states into transients and non-transients of $M_0$*}
9:      $M_\epsilon$.scale($N$)
10:      $s$.zeroColumns($T$)
11:    **else**
12:      $(p_0, i_0^*) = $ quotient($M_0, Q$) {*compute a normalized quotient*}
13:      $M_\epsilon = $ dropHighOrderTerms($p_0 (M_\epsilon - I) i_0^* + I$)
14:      $M_\epsilon$.scale()
15:      $s = si_0^*$
16:    **end if**
17: **end while**
18: $Q = $ getStatesExceptClosedClassReps($M_0$)
19: $(p_0, i_0^*) = $ quotient($M_0, S$) {*compute a normalized quotient, which eliminates any remaining transient states*}
20: **return** $si_0^*$

ble distribution of a perturbed Markov process (see Algorithm 1). Steps 10 and 20 guarantee that transient states are not in the support of the stochastically stable distribution. Thus, with this algorithm, we can conveniently identify the stochastically stable states of a perturbed Markov process. If we define the set of *eventually transient* states as those states that become transient in some iteration of the algorithm, then:

**Corollary 5.7** *If $M_\epsilon$ is a perturbed Markov process, then the stochastically stable states of $M_\epsilon$ are precisely those states which are* not *eventually transient.*

## 6 Discussion

Previously, the only known "algorithm," to our knowledge, for computing a stochastically stable distribution was an iterative approach that falls out of the definition: e.g., compute the stable distribution of $M_{0.1}$, compute the stable distribution of $M_{0.01}$, compute the stable distribution of $M_{0.001}$, etc., and compute the limit of this sequence of stable distributions. However, since the computation of such distributions can be numerically unstable, such a sequence may appear to be converging to the *wrong* limit, or the limit may appear to be undefined.

For example, consider the Markov process

$$M_\epsilon = \begin{pmatrix} 0.5 & 0.5 & 0 & 0 & \epsilon^3 \\ 0.5 & 0.5 - \epsilon^5 & 0 & 0 & 0 \\ 0 & \epsilon^5 & 0.5 & 0.5 & 0.5 \\ 0 & 0 & 0.5 & 0.5 - \epsilon^2 & 0 \\ 0 & 0 & 0 & \epsilon^2 & 0.5 - \epsilon^3 \end{pmatrix} \quad (13)$$

When $\epsilon = 0.001$, *Mathematica* computes a unique stable distribution of (approximately) $\begin{pmatrix} 0.37 & 0.37 & 0.13 & 0.13 & 2.6 \times 10^{-7} \end{pmatrix}^t$, while Matlab gives $\begin{pmatrix} 0.31 & 0.31 & 0.19 & 0.19 & 3.9 \times 10^{-7} \end{pmatrix}^t$. However, when $\epsilon \geq 10^{-4}$, *Mathematica* suggests that $M_\epsilon$ has a unique stable distribution of $\begin{pmatrix} 0.5 & 0.5 & 0 & 0 & 0 \end{pmatrix}^t$, while Matlab implies that $M_\epsilon$ has two, independent eigenvectors, leading to an infinite number of possible distributions. The unique stable distribution is actually $\begin{pmatrix} 2+4\epsilon^5 & 2 & 1+2\epsilon^2+2\epsilon^3+4\epsilon^5 & 1+2\epsilon^3 & 2\epsilon^2 \end{pmatrix}^t / (6 + 4\epsilon^2 + 4\epsilon^3 + 8\epsilon^5)$. Due to the small values in the $(3,2)$- and $(5,4)$-entries, magnification of round-off error makes it difficult to obtain the true stochastically stable distribution, $\begin{pmatrix} \frac{1}{3} & \frac{1}{3} & \frac{1}{6} & \frac{1}{6} & 0 \end{pmatrix}^t$.

# 7 Adaptive Learning in Coordination Games

In this section, we compute the stochastically stable distribution of Young's model of adaptive learning in coordination games with multiple equilibria.

In adaptive learning, each player maintains a finite history of its opponents' $m$ most recent moves. For example, in the QWERTY game, if $m = 4$, the state of the game is given by two sequences, say, $dqqd$ for player 1, indicating player 2's history of play, and $DQDQ$ for player 2, indicating player 1's history of play. Concatenating these histories yields the states of the underlying Markov process. If $m \geq 4$, there are 256 states.

Although the state includes the history of play for the past $m$ moves, in the adaptive learning model, each player has only limited information about the state of the game. Specifically, each player has access to a random sample of size $s$ from its opponents' history, and must make a decision based on the distribution of actions in that sample. For example, if $s = 2$ and the state is $dqqdDQDQ$, then there is a $\frac{1}{6}$ chance that player 1 draws the sample $\{q,q\}$. This suggests to player 1 that, in recent history, player 2 has always played $q$. The samples presented to each player are chosen independently. For example, there is a $\frac{4}{36}$ chance of player 1 drawing the sample $\{q,q\}$ while player 2 draws the sample $\{D,Q\}$.

Finally, in adaptive play, it is assumed that each player has limited rationality. That is, each player usually plays a best-response to his perception of his opponents' history of play (in particular, the distribution of play, as in fictitious play). Specifically, he will choose a best-response action, with probability $1 - \epsilon$ (making a uniformly random choice among equally valuable options), or a uniformly random choice among all possible options, with probability $\epsilon$. His choice depends on his own history of play only in so much as his actions help to shape the actions of his opponents.

**Game 2: Coordination Game**

|   | a   | b   |
|---|-----|-----|
| A | 5,5 | 0,3 |
| B | 3,0 | 4,4 |

If $m = s = 1$, the following irreducible Markov matrix expresses the adaptive learning model's dynamics in the QWERTY game:

$$M_\epsilon = \frac{1}{4} \begin{pmatrix} 4-4\epsilon+\epsilon^2 & 2\epsilon-\epsilon^2 & 2\epsilon-\epsilon^2 & \epsilon^2 \\ 2\epsilon-\epsilon^2 & \epsilon^2 & 4-4\epsilon+\epsilon^2 & 2\epsilon-\epsilon^2 \\ 2\epsilon-\epsilon^2 & 4-4\epsilon+\epsilon^2 & \epsilon^2 & 2\epsilon-\epsilon^2 \\ \epsilon^2 & 2\epsilon-\epsilon^2 & 2\epsilon-\epsilon^2 & 4-4\epsilon+\epsilon^2 \end{pmatrix}$$

After one iteration of our algorithm, we are left with the following matrix, which is regular even when $\epsilon = 0$:

$$M_\epsilon = \frac{1}{4} \begin{pmatrix} 2-\epsilon & 2 & \epsilon \\ 2 & 0 & 2 \\ \epsilon & 2 & 2-\epsilon \end{pmatrix}$$

Here, the first and last states correspond to the stable distributions on $dD$ and $qQ$, respectively, while the second state corresponds to the uniform distribution on states $dQ$ and $qD$. Since the stable distribution of this system is $\begin{pmatrix} \frac{1}{4} & \frac{1}{2} & \frac{1}{4} \end{pmatrix}^T$, the stochastically stable distribution of the original system is the uniform distribution over all four states.

QWERTY is a typical example of a coordination game. Another interesting coordination game is Game 2. Young argues that state $b \cdots bB \cdots B$, the so-called "risk-dominant convention", is the unique stochastically stable state in all such games, for $s$ and $m$ sufficiently large and $\frac{s}{m} \leq 0.5$ (Young, c1998). Using our algorithm, we quickly determined the stochastically stable distribution for $1 \leq m, s \leq 4$. We found that $b \cdots bB \cdots B$ is the unique stochastically stable state whenever $s \geq 2$. This suggests that Young's result could possibly be strengthed to eliminate the restriction that $\frac{s}{m} \leq 0.5$.

Battle of the Sexes (Game 3) is another coordination game. Here, a man and a woman would like to attend the same event, either the ballet or the football game. The woman prefers the ballet, while the man prefers the football game, but both prefer to be attend the same, rather than different, events.

As above, we computed the stochastically stable distribution for $1 \leq m, s \leq 4$ and found that in almost every case (with the exceptions of $(m,s) = (1,1), (2,2)$), the stochastically stable distribution was uniform on the two states $b \cdots bB \cdots B$ and $f \cdots fF \cdots F$. In other words, almost all the time the players either agreed to both go to the ballet or both go to the football

**Game 3: Battle of the Sexes**

|   | b | f |
|---|---|---|
| B | 2,1 | 0,0 |
| F | 0,0 | 1,2 |

**Game 4: Biased Battle of the Sexes**

|   | b | f |
|---|---|---|
| B | 3,1 | 0,0 |
| F | 0,0 | 1,2 |

game for an extended period of time, with a negligible amount of time spent transitioning between these two conventions or exploring other possibilities, and an equal amount of time spent in each convention.

As might be expected, if the payoffs are adjusted to favor the ballet, as in Game 4 the players select this convention. In particular, for $s \geq 3$, $b \cdots bB \cdots B$ is the unique stochastically stable state.

It is interesting to note that the stochastically stable distribution is sensitive to dependencies between the payoff matrices and the parameter values. For example, in Battle of the Sexes when $(m, s) = (2, 2)$, the stochastically stable distribution is given by Table 1. On average, players spend an equal amount of time playing one of the following three conventions: both go to the ballet, both go to the football game, or play cycles from $bbFF$ to $bfFB$ to $fbBF$ to $bbFF$, and so on. The third convention may be described briefly as follows: if the opponent has made the same play twice in a row, make the corresponding play; otherwise, player 1 plays $B$ and player 2 plays $f$.

Table 1: Battle of the Sexes, $(m, s) = (2, 2)$.

| bbBB | bbFF | bfFB | fbBF | ffFF |
|---|---|---|---|---|
| 1/3 | 1/9 | 1/9 | 1/9 | 1/3 |

## 8  Summary and Future Work

In summary, we have presented the first exact algorithm, to our knowledge, for computing the stochastically stable distribution of a perturbed Markov process. We have used our algorithm to explore an adaptive learning model in sample repeated games. We verified that the notion of stochastic stability does indeed offer at least a partial solution to the equilibrium selection problem for multiagent learning in repeated games. In future work, we are interested in studying the dynamics of other learning models, such as regret-matching (Hart and Mas-Colell, 2000). We expect that our technology will prove useful to AI researchers who are building multiagent systems, since it will enable them to better predict the long-term dynamics of multiagent learning environments. We also expect our technology to prove useful to AI researchers, more generally. We are presenting exploring applications of this work to other areas of AI, ranging from simulated annealing to spectral clustering to information retrieval.